\documentclass[aps,prl,twocolumn,showpacs, amssymb, amsmath, aps, superscriptaddress] {revtex4-1}

\usepackage{amsmath,amssymb, bbold, bbm, soul,bm}
\usepackage[pdftex]{graphicx}
\usepackage{epstopdf}
\usepackage{color}
\usepackage{algorithm}

\newcommand{\ket}[1]{| #1 \rangle}
\newcommand{\bra}[1]{\langle #1 |}

\newcommand{\ketbra}[2]{|#1 \rangle\!\langle #2 |}
\newcommand{\matele}[3]{\langle #1 | #2 | #3 \rangle}

\newcommand{\tr}{{\rm Tr}}

\begin{document}

\title{Experimental study of optimal measurements for quantum state tomography}

\author{H. Sosa-Martinez}\affiliation{Center for Quantum Information and Control, College of Optical Sciences and Department of Physics, University of Arizona, Tucson, AZ 85721, USA}
\author{N. K. Lysne}\affiliation{Center for Quantum Information and Control, College of Optical Sciences and Department of Physics, University of Arizona, Tucson, AZ 85721, USA}
\author{C. H. Baldwin}\affiliation{Center for Quantum Information and Control, Department of Physics and Astronomy, University of New Mexico, Albuquerque, NM 87131, USA}
\author{A. Kalev}\affiliation{Center for Quantum Information and Control, Department of Physics and Astronomy, University of New Mexico, Albuquerque, NM 87131, USA}
\author{I. H. Deutsch}\affiliation{Center for Quantum Information and Control, Department of Physics and Astronomy, University of New Mexico, Albuquerque, NM 87131, USA}
\author{P. S. Jessen}\affiliation{Center for Quantum Information and Control, College of Optical Sciences and Department of Physics, University of Arizona, Tucson, AZ 85721, USA}

\begin{abstract}
Quantum tomography is a critically important tool to evaluate quantum hardware, making it essential to develop optimized measurement strategies that are both accurate and efficient. We compare a variety of strategies using nearly pure test states. Those that are informationally complete for all states are found to be accurate and reliable even in the presence of errors in the measurements themselves, while those designed to be complete only for pure states are far more efficient but highly sensitive to such errors. Our results highlight the unavoidable tradeoffs inherent to quantum tomography.
\end{abstract}

\maketitle

Progress in quantum information science has now reached the point where rudimentary quantum computers are appearing in the laboratory~\cite{Barends14, Corcoles15, Riste15, Ofek16, Takita16, Debnath16, Monz16, Linke17}. As these devices grow in complexity it becomes more difficult to verify that their building blocks perform as required and to identify physical sources of error so they can be countered. In this situation quantum tomography seems an ideal diagnostic tool, capable of providing complete estimates of quantum states~\cite{Vogel89}, processes~\cite{Chuang97}, and measurements~\cite{Luis99}.  In practice, however, its use has been limited by its own inherent challenges: tomography is based on measurement data and the process of collecting it is itself subject to error.  As a result there has been great interest in optimized measurement strategies that make tomography as efficient and accurate as possible. 

In this letter we present results from a comprehensive experimental study of different measurement strategies for Quantum State Tomography (QST), each corresponding to a particular choice of POVM.  The notion of an optimal POVM is nuanced and context dependent.  For example, two different POVMs, the Symmetric Informationally Complete (SIC) POVM~\cite{Renes04} and Mutually Unbiased Bases (MUB)~\cite{Wootters89}, provide optimally {\it accurate} reconstruction on average~\cite{Scott06}, but the theoretical assumptions for this to be true may not hold in the laboratory. Other notions of optimality arise when one employs prior information. Of particular interest are POVMs that are optimally {\it efficient}, requiring a minimal number of measurement outcomes given some prior knowledge about the state, e.g., that it is close to pure~\cite{Flammia05,Chen13,Goyeneche14,Finkelstein04,Carmeli14,Carmeli15,Carmeli16,Kalev15,Baldwin16}.  While efficient, these strategies can be compromised by experimental imperfection in ways that are usually not included in theoretical analyses. 

So far, a number of experiments have provided proof-of-principle demonstrations of various measurement strategies for QST~\cite{Ling06,Adamson10,Medendrop11,Lima11,Giovannini13,Pimenta13,Bent15},  but the diversity of experimental platforms has made it difficult to compare their performance. As a result, one of the central questions is still to be addressed: What are the relative merits of different POVMs in a real-world scenario where the assumptions underlying optimality and/or priors may or may not apply?  We address that question by implementing a comprehensive collection of POVMs and using them for QST on a common physical platform. This test bed consists of the $d=16$ dimensional Hilbert space formed by the coupled electron-nuclear spins of individual $^{133}$Cs atoms in the electronic ground state~\cite{Smith13, Anderson15}. As expected for a well-behaved system, we find that accurate and efficient QST can generally be achieved across a large sample of arbitrarily chosen nearly-pure test states.  More significantly, our results provide new insight into the tradeoffs between efficiency, accuracy, and robustness inherent to different POVMs. 

A key concept for QST is that of an Informationally Complete (IC) POVM.  A Fully-IC  POVM allows one to identify an arbitrary unknown density matrix from measurement data (in the absence of noise and errors).  The most efficient Fully-IC POVM is the SIC-POVM which has the minimal number of POVM outcomes, $d^2$~\cite{Renes04}.  Other examples of Fully-IC POVMs include the $d+1$ Mutually Unbiased Bases (MUB) ~\cite{Wootters89}, and the $2d-1$ generalized Gell-Mann Bases (GMB) ~\cite{Baldwin16}, with $d^2+d$ and $2d^2-d$ outcomes, respectively.  Additional notions of IC become relevant for QST on restricted subsets of states. Notably, quantum information processing tends to rely on pure states, and diagnostic tools such as randomized benchmarking~\cite{Knill08,Magesan12} can verify that a given experiment operates close to this regime.  We  thus test several efficient strategies for QST of rank-1 density operators:  Rank-1 IC (R1-IC) POVMs, which uniquely identify a pure state {\em only} from other pure states~\cite{Flammia05,Carmeli14,Baldwin16}, and Rank-1 {\em Strictly} IC (R1S-IC) POVMs, which uniquely identify a pure state from {\em all} physical density matrices of any rank~\cite{Chen13,Carmeli14,Baldwin16}.  Strictly IC POVMs are particularly useful because they allow accurate state estimation via convex optimization~\cite{Baldwin16}.  

The notion of efficient tomography given prior information is  related to compressed sensing tomography~\cite{Gross10, Flammia12}, with some subtle differences.  The compressed sensing protocol involves a specific class of measurements and a specific form of convex optimization ~\cite{Candes10,Recht10}.  While we have shown that all compressed sensing measurements for pure states are R1S-IC~\cite{Kalev15}, the converse is not true. Nevertheless, as we will see here, QST with a R1S-IC POVM is similar to compressed sensing insofar as the two protocols achieve high accuracy from similar amounts of data, in both cases much less than required for a Fully-IC POVM.

Our experiments explore a variety of POVM constructions that have been studied in the literature.  Flammia~{\em et al.} introduced a R1-IC POVM that contains $3d-2$ elements, each of which are rank-1 (nonprojective) operators~\cite{Flammia05}.  We refer to this POVM as PSI, and it can be shown to be R1S-IC by the method proposed in~\cite{Baldwin16}. PSI has the best known scaling with $d$ of any R1S-IC POVM that consists of rank-1 elements. We implement two additional R1S-IC POVMs: the first of which we refer to as 5 Gell-Mann Bases (5GMB), originally proposed in~\cite{Goyeneche14} and proven to be R1S-IC in~\cite{Baldwin16}, and the second of which we refer to as the 5 Polynomial Bases (5PB)~\cite{Carmeli16}.  We further implement two R1-IC POVMs, which we refer to as 4 Gell-Mann Bases (4GMB)~\cite{Goyeneche14} and 4 Polynomial Bases (4PB)~\cite{Carmeli15}. The minimum number of orthonormal bases needed to reconstruct a pure state is 4 (for $d \geq 5$)~\cite{Carmeli15}; such POVMs are R1-IC but not R1S-IC. Details on how we construct the various POVMs can be found in~\cite{CharlieThesis, HectorThesis}.

Our experimental test bed has been described elsewhere~\cite{Smith13,Anderson15} and only the most important features are summarized here.  A $^{133}$Cs atom in the $6S_{1/2}$ electronic ground state has electron and nuclear spins $S=1/2$ and $I=7/2$, resulting in two hyperfine manifolds with spin $F=I \pm S = 3,4$ and a combined  total of $16$ magnetic sublevels $\ket{F,m}$. The system is controllable in this $d=16$ dimensional Hilbert space $\mathcal{H}$ with a combination of a static magnetic field and phase modulated radio-frequency and microwave magnetic fields. The phase modulation (control) waveforms used to implement a given unitary are found though numerical optimization; these control waveforms are not unique and if desired it is straightforward to find several high-performing ones.  In previous experiments we have verified through randomized benchmarking that we can implement a variety of control tasks with high accuracy,  ranging all the way from arbitrary quantum state-to-state maps (average infidelity 0.005(1)) to arbitrary SU(16) maps (average infidelity 0.018(2)).

A typical experimental sequence begins with an ensemble of  $\sim 10^6$ laser cooled atoms released into free fall.  We use optical pumping to prepare the ensemble in $\ket{\psi_0}=\ket{F=3,m=3}$ and then implement a state map $\ket{\psi_0}\rightarrow\ket{\psi_{t}}=\sum_{F,m} c_{F,m} \ket{F,m}$ to obtain a desired test state. In practice, finite errors and  imperfections in the preparation sequence cause the actual state, $\rho_a$,  to deviate slightly from the intended target, with an average infidelity $\bra{\psi_t}\rho_a\ket{\psi_t}\approx0.005$.  This is our starting point for QST. 

\begin{figure}
[t]\resizebox{8.25cm}{!}
{\includegraphics{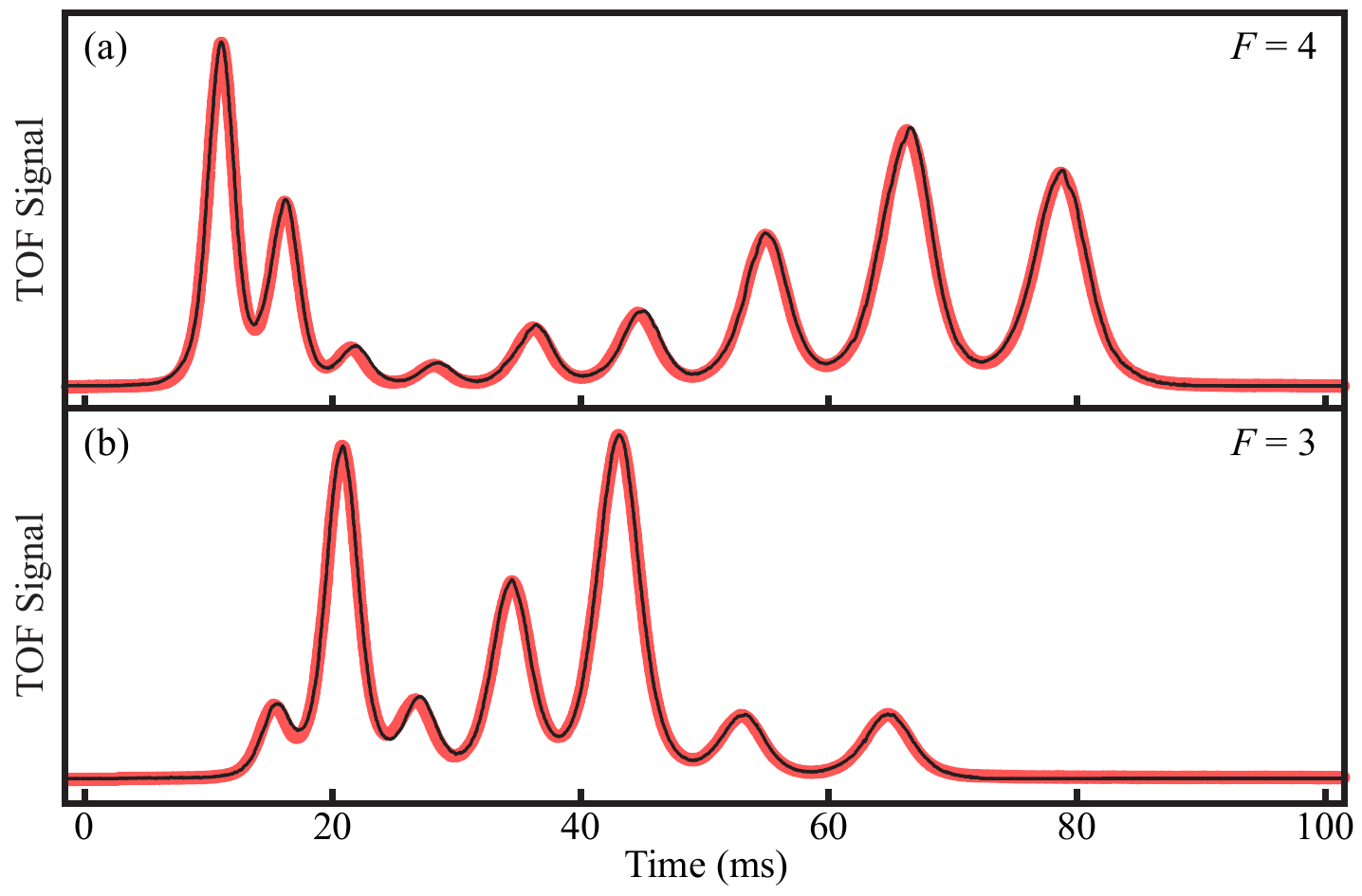}}
\caption{\label{fig:Fig1} (Color online) Stern-Gerlach analysis of  $^{133}$Cs atoms in a hyperfine state with support on all 16 magnetic sublevels.  (a) Time-of-Flight signal from atoms in the $F=4$ manifold.(b) Same from atoms in the $F=3$ manifold. Black lines are measured signals, light red lines are fits.}
\end{figure}

The basic resource for measurement in our experiment is Stern-Gerlach analysis, implemented by letting the atoms fall in a magnetic field gradient until they reach an optical probe a few cm below the preparation volume~\cite{Klose01}. Fig. 1 shows a typical Time-Of-Flight (TOF) signal for a state with support on all 16 magnetic sublevels. Given the partial overlap between the arrival distributions for some $\ket{F,m}$ such a signal is not by itself a projective measurement.  Nevertheless, we can fit it to a weighted sum of arrival distributions, $S_{TOF}(t) = \sum_{F,m} \nu_{F,m} S_{F,m} (t)$, and from those weights get a good estimate of the probability of each outcome,  $\nu_{F,m} \approx  p_{F,M} = \bra{F,m}\rho_a \ket{F,m}$. This is functionally equivalent to finding the frequency of outcomes $\nu_{F,m}$ from a series of separate, projective measurements on the individual atoms in the ensemble.  A significant advantage of working with $\sim 10^6$ atoms in parallel is that it speeds up data acquisition, to the point that measurement statistics are effectively eliminated as a source of error.  Fluctuations in the probe power and electronic noise in the detector contribute a roughly $1\%$ statistical uncertainty in the estimated frequencies.

With the basic Stern-Gerlach measurement in place, our ability to perform SU(16) maps in $\mathcal{H}$  makes it straightforward to implement more general POVMs.  Specifically, we can perform a 16-outcome measurement in an arbitrary orthonormal basis $\{\ket{\phi_\mu}\}$ in  $\mathcal{H}$, by mapping each $\ket{\phi_\mu}$ onto a magnetic sublevel $\ket{(F,m)_\mu}$ through a unitary transformation $U=\sum_{\mu}\ket{(F,m)_\mu}\bra{\phi_\mu}$. An arbitrary POVM consisting of a collection of different orthogonal bases, e.g., MUBs, can be constructed from a series of such measurements in different runs of the experiment, yielding POVM elements, $\{ E^{(i)}_\mu = \ketbra{\phi^{(i)}_\mu}{\phi^{(i)}_\mu} \}$, where $i=1,2,\dots,n$ labels the basis.  Furthermore, by restricting the test state to a subspace $\mathcal{H}'\subset \mathcal{H}$, we can use the Neumark extension~\cite{Peres1995} to implement nonorthogonal POVMs with up to 16 outcomes, each specified by a rank-1 POVM element. To do so, all that is required is an orthogonal measurement in $\mathcal{H}$ chosen such that the projection of its measurement operators onto $\mathcal{H}'$  yields the desired POVM elements, $\tilde{E}^{(i)}_\mu = \Pi E^{(i)}_\mu \Pi^\dagger$. 

The above approach to measurement has one very important consequence: the performance of QST is dominated by systematic errors in the POVMs rather than statistical noise from working with a finite number of copies of the state.  Our unitary maps are subject to a roughly $2\%$ error arising mainly from a fixed inhomogeneity of the control Hamiltonian across the ensemble, and repeated implementations of a given map with the same control waveform therefore results in the same fixed measurement error. At the same time, different unitary maps, or even the same map implemented with different control waveforms, will have systematic errors that are largely uncorrelated. We note that errors of this type are present (if not necessarily dominant) in most implementations of QST regardless of the physical platform at hand, since detectors are generally designed to measure in a fixed basis and additional POVMs are performed by preceding the measurement with unitary maps. 

The final element of QST is the data processing algorithm used to obtain a state estimate.  Standard estimators include maximum-likelihood~\cite{Hradil97},  least-squares~\cite{James01}, and more recently, trace-minimization~\cite{Gross10}.  Because our focus is on the relative performance of the various POVMs, it suffices to pick one estimator and apply it consistently across all our data sets.  We chose here the maximum-likelihood estimator (MLE): 
\begin{align}
&\hat{\rho} =\arg\!\min_\rho\;\;  -\sum_{j} \nu_j \log p_j,\text{ s.t.: }  \tr[\rho] =1,  \; \rho\geq 0, \nonumber
\label{eq:eq1}
\end{align}
\noindent where the summation is over experimental outcomes. Efficient solution of this numerical optimization problem is achieved by convex optimization using the Matlab package CVX~\cite{cvx}.

\begin{table}
\begin{ruledtabular}
\begin{tabular}{llcc}
\ & \ & \multicolumn{2}{c}{Average infidelity ($\Delta$)}\\
IC class&POVM&$\it{d}$=4&$\it{d}$=16\\
\hline
Fully-IC  	& SIC & 0.0625 (73) & N/A \\
              	& MUB & 0.0181 (21) & 0.0602 (23) \\
 		& GMB & 0.0092 (15) & 0.0595 (30) \\
\hline
R1S-IC  	& PSI & 0.0923 (164) & N/A \\
              	& 5MUB & N/A & 0.1564 (96) \\
 		& 5GMB & 0.0173 (28) & 0.2442 (173) \\
		& 5PB & 0.0267 (47) & 0.2384 (215) \\
\hline
R1-IC  	& 4GMB & 0.0764 (221) & 0.2759 (217) \\
              	& 4PB & 0.0853 (360) & 0.3200 (366) \\
\end{tabular}
\end{ruledtabular}
\caption{Average QST infidelities $\Delta$ achieved for nine different POVMs in three different IC categories, in Hilbert spaces with dimensions $d=4$ and $d=16$.  Parentheses indicate uncertainties of one standard deviation.}
\end{table}

Our experimental study covers 8 POVM constructions (SIC, MUB, GMB, PSI, 5GMB, 5PB, 4GMB, 4PB) as described above, applied in a randomly chosen subspace  $\mathcal{H}'$ ($d=4$); this is the largest subspace for which we can apply the Neumark extension to implement the 16-outcome SIC-POVM as well as the 10-outcome PSI-POVM.  The remaining 6 POVMs are also applied to states in the full Hilbert space, $\mathcal{H}$ ($d=16$) along with a POVM consisting of 5 MUBs that is expected to be R1S-IC~\cite{CharlieThesis}.  In each case the performance of QST was evaluated by applying the given protocol to a set of 20 randomly chosen test states, $\{\ket{\psi^{(j)}_t }\}$, and calculating for each the infidelity between the input and the MLE reconstruction, $\Delta_j = 1 - \matele{\psi^{(j)}_t }{\hat{\rho}}{\psi^{(j)}_t} $. 

Table 1 and Fig. 2 show the mean and variance of the set $\{\Delta_j\}$ observed for each POVM.  Several immediate observations can be made from this data.  First, the average error (infidelity) of the estimates varies considerably with measurement strategy, ranging from 0.02(2) to 0.10(15) in $d=4$, and from 0.06(3) to 0.27(2) in $d=16$. Second, there is a very clear tradeoff between the accuracy and efficiency of QST: the Fully-IC POVMs tend to perform better than R1S-IC POVMs, which in turn tend to perform better than R1-IC POVMs.  Finally, SIC and PSI perform significantly worse than other POVMs from within the same IC class.

\begin{figure}
[t]\resizebox{8.25 cm}{!}
{\includegraphics{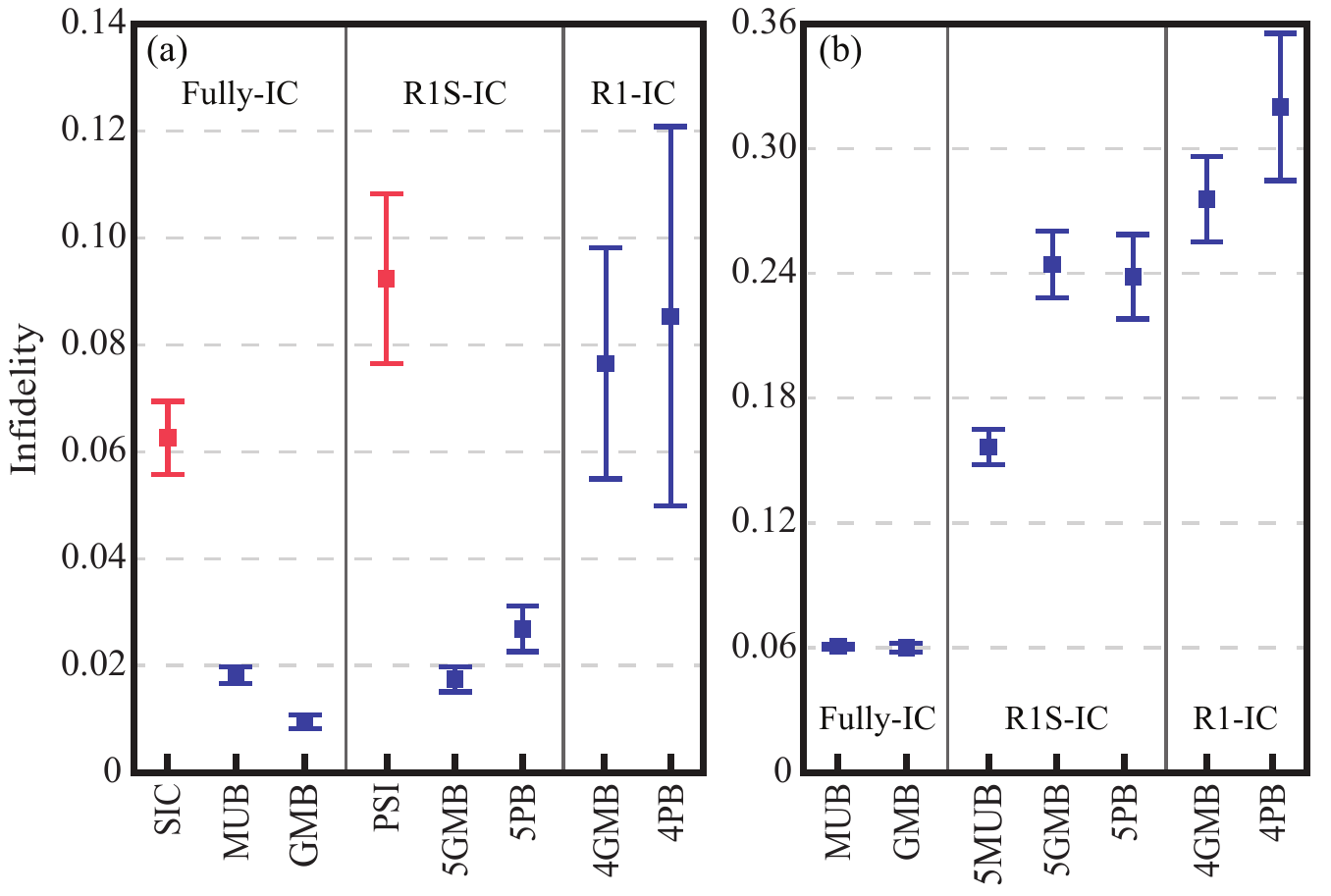}}
\caption{\label{fig:Fig2} (Color online) Average QST infidelities $\Delta$ for nine POVMs in three IC categories (data fromTable 1). (a) QST in $d=4$. (b) QST in $d=16$. POVMs consisting of multiple orthogonal measurements are shown in (dark) blue, POVMs consisting of a single nonorthogonal measurement with $>d$ outcomes are shown in (light) red}
\end{figure}

The difference in QST performance achieved with different POVMs can be ascribed to a variety of factors.  Most importantly, the fact that systematic errors dominate in our experiment will favor POVMs such as MUB and GMB that use many orthogonal measurements with many different outcomes.  The reason is two-fold. First, when applied to pure states specified by $2d-2$ real-valued parameters, the large number of outcomes provide redundant information. And second, each orthogonal measurement uses a unitary map with its own distinct errors, in which case a larger number of measurements provide better averaging over experimental imperfections and improve the accuracy of QST.  We believe this largely accounts for the superior performance of Fully-IC POVMs relative to the R1S-IC POVMs (5GMB, 5PB).

Similar considerations play out for the nonorthogonal POVMs in $d=4$.  SIC is Fully-IC and its $d^2=16$ outcomes are redundant for pure states, whereas PSI is R1S-IC and slightly less redundant with $3d-2=10$ measurement outcomes.  As one might expect on this basis, SIC performs somewhat better than PSI.  At the same time, because each is implemented with a single unitary/Stern-Gerlach measurement there is no averaging over errors in the POVMs, and both perform significantly worse than POVMs that measure several orthogonal bases.  The comparison between SIC and MUB is especially instructive: these POVMs are equally optimal when QST is limited solely by measurement statistics~\cite{Scott06}, but MUB performs much better when systematic errors dominate. 

While averaging over systematic errors in the POVMs explains much of the variation we observe in QST performance, other considerations also come into play. The number of orthogonal measurements does not differ greatly between R1S-IC POVMs and R1-IC POVMs. However, R1-IC POVMs can correctly identify the state only from within the restricted set of pure states.  This is a problem in our protocol because the measurement record may be better matched by a distant mixed state than by the actual, nearly pure state present in the experiment. In that situation the MLE algorithm will identify the distant state as the best estimate.  As seen in Fig. 2, this leads to a significant increase in average infidelity. 

\begin{figure}
[t]\resizebox{8.25cm}{!}
{\includegraphics{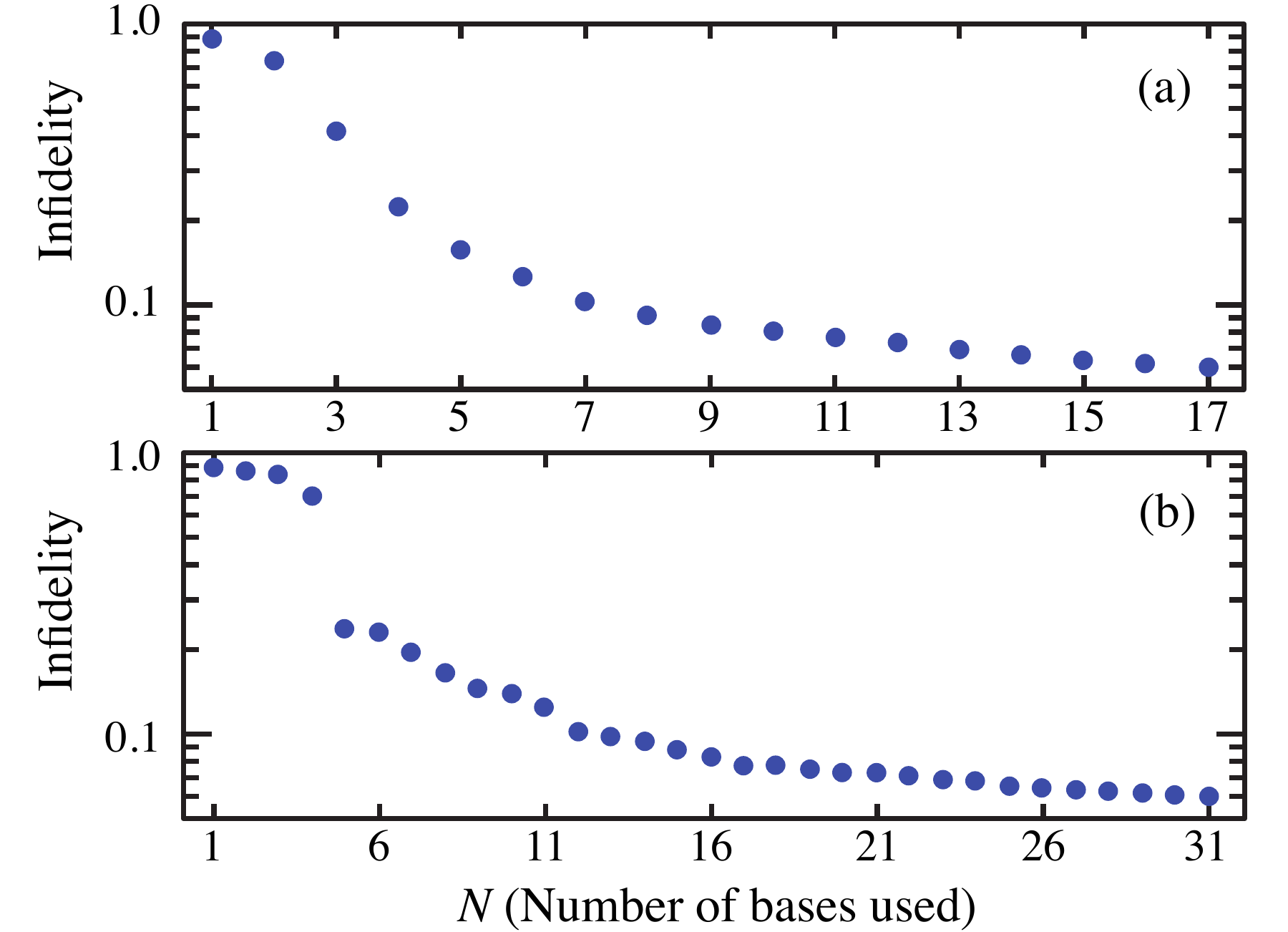}}
\caption{\label{fig:Fig3} QST of nearly-pure test states in $d=16$, using data from POVMs that are less than Fully-IC.  (a) Infidelity versus the number $N$ of MUBs used (Fully-IC for $N=17$).  (b) Infidelity versus number of GMBs used (Fully-IC for $N=31$).  Both POVMs become Strictly-IC at $N=5$ }
\end{figure}

Yet another issue in the performance of R1-IC and R1S-IC POVMs is the possibility of ``failure sets." Some of these POVMs are designed to give an analytic relationship between the state and the probabilities associated with the measurement outcomes, e.g. the relationship shown in~\cite{Flammia05}.  This inversion fails on a known set of measure zero, which corresponds to states that have zero probability of certain outcomes.  In practice, the presence of noise and errors will extend the failure set to a finite measure, and there will be a finite probability that a randomly chosen state will fall within it~\cite{Finkelstein04}.  Among the POVMs examined here, only PSI and 5GMB have failure sets.  Comparing 5GMB against 5PB we see no significant difference in performance, indicating that the presence or absence of a failure set has little effect.  As already discussed, PSI performs poorly for other reasons and it is difficult to isolate the effect of its failure set.  Overall, the consequence of failure sets is inconclusive in our data set, and further work will be necessary to understand how they affect QST in the presence of noise and errors.

A final, vivid illustration of the tradeoff between efficiency and robustness can be had by considering POVMs that fall between R1S-IC and Fully-IC POVMs.  Figure 3 shows how the average infidelity of QST in $d=16$ improves when using an increasing number of the bases, $N$, making up the MUB and GMB POVMs. In both cases we see a clear ``compressed-sensing effect" with an infidelity that drops rapidly until the POVM becomes R1S-IC at $N=5$, and then slowly improves as additional measurements provide redundancy and help average out systematic errors.  Measuring more bases improves performance, but for an application that can tolerate a certain level of infidelity it may be preferable to use something well short of a Fully-IC POVM with its large data-taking overhead.

Looking ahead, there are several important aspects of QST that might be addressed.  Most importantly, the use of QST as a diagnostic tool puts a premium on its accuracy and on systematic ways to improve it. Consider the fairly typical scenario exemplified by our experiment, in which the average fidelity of state preparation is significantly better than that of QST.  In principle this means we can use known input states to perform POVM tomography~\cite{Luis99}, which should allow us to correct or account for systematic errors in our POVMs and thereby improve our QST fidelity--which in turn might allow us to further optimize our state preparation. This suggests the possibility of a virtuous cycle of successive improvements to state preparation, POVM implementation, and QST. Other questions pertain to the role of estimators in QST. It is possible in principle to combine R1-IC POVMs with an estimator that searches only over pure states; preliminary results from our experiment suggests this approach can be very successful at diagnosing state preparation errors that are coherent in nature, while at the same time largely ignoring the presence of other types of error~\cite{CharlieThesis}. This in turn raises the issue of bias in QST, which has been previously discussed in the context of compressed sensing approaches~\cite{Riofrio17}. Such questions might be studied in experiments using known mixed test states, a prospect within the current capabilities of our cold-atom test bed. 

\begin{acknowledgments}
This work was supported by the U. S. National Science Foundation Grants No. PHY-1521439, No. PHY-1521431, and No. PHY 1521016.
\end{acknowledgments}

\end{document}